\documentstyle[preprint,aps]{revtex}
\begin{document}
\draft
\title{Universal Scaling Properties in Large Assemblies of Simple
Dynamical Units\\
Driven by Long-Wave Random Forcing}
\author{Yoshiki Kuramoto and Hiroya Nakao}
\address{Department of Physics, Graduate School of Sciences,
Kyoto University, Kyoto 606, Japan}
\maketitle
\begin{abstract}
Large assemblies of nonlinear dynamical units 
driven by a long-wave fluctuating
external field are found to generate 
strong turbulence with scaling properties. 
This type of turbulence is so robust that it persists 
over a finite parameter range with parameter-dependent 
exponents of singularity, 
and is insensitive to the specific nature of the
dynamical units involved.
Whether or not the units are coupled with their neighborhood
is also unimportant.
It is discovered numerically that the derivative of the
field exhibits strong spatial intermittency with multifractal structure.
\end{abstract}
\narrowtext
We report our discovery of a new type of turbulent behavior
which arises generally
in large assemblies of simple dynamical units 
driven by a long-wave randomly fluctuating field. 
The driving field may 
actually be a self-generated internal field due to long-range interaction, 
and this particular situation was studied in a previous paper [1] 
where a rough explanation of the origin of 
power-law correlations was also given. The present study thus aims 
at expanding as much as possible the class of systems capable of 
exhibiting the same type of turbulence, 
and also proposing a more transparent 
and coherent explanation of the phenomena.\\
\indent
An illustrative example is given by an array of {\em uncoupled} logistic 
maps $f(X)=aX(1-X)$ with driving 
%Eq.(1)
\begin{equation}
X_{n+1}(j)=f(X_{n}(j))+h_{n}(j),\;\;\;j=1,2,\ldots N,
\end{equation} 
where $h_n(j)=\frac{K}{2}
\left(1+\cos[2\pi\{\frac{j}{N}+\psi_n\}]\right)$, $\psi_n$ being a 
random variable in the interval $(0,1]$ with uniform distribution. 
Were it not for spatial dependence of $h_n$,
Eq.\ (1) would represent $N$ identical copies 
of a randomly driven map; the dynamics of such an ensemble was
studied in [2]. Making $h_n$ nonuniform 
changes the
problem completely. 
Let the parameter values be set such that individual maps are 
entrained to $h_n$ in the sense that their maximum Lyapunov 
exponent (common to all maps) is negative. 
Unlike $h_n$, however, the corresponding amplitude 
profile is not smooth at all, a typical example 
of which is displayed in Fig.\ 1a. Such ill-behaved nature of the 
pattern is even amplified in Fig.\ 1b which shows a strongly 
intermittent pattern of the {\em differential amplitudes} 
$Y(j)\equiv|X(j+1)-X(j)|$ constructed from Fig.\ 1a. \\
\indent Before proceeding to further numerical study, some theoretical
predictions will be made as to the statistics of turbulence to be shared
by the above system or more general assemblies of units 
under long-wave random 
driving. For this purpose, it seems more convenient 
to work with a picture in which the dynamical units form 
a quasi-continuum rather than a lattice, and  
the driving field has a characteristic wavelength of $O(1)$. 
Our primary concern is to understand how a simple driving field
alone can generate a 
nontrivial correlation between uncoupled units. \\
\indent 
Suppose that each unit is represented by a discrete-time 
dynamical system 
$\mbox{\boldmath$X$}_{n+1}=\mbox{\boldmath$f$}(\mbox{\boldmath$X$}_n)$.
Let the units be driven by an additive random force $\mbox{\boldmath$h$}_n$
which is smooth in space and statistically invariant
with respect to spatial translations.  
The unit 
at site $\mbox{\boldmath$r$}$ is governed by the equation
%eq.(2)
\begin{equation}
\mbox{\boldmath$X$}_{n+1}(\mbox{\boldmath$r$})=
\mbox{\boldmath$f$}(\mbox{\boldmath$X$}_n(\mbox{\boldmath$r$}))
+\mbox{\boldmath$h$}_n(\mbox{\boldmath$r$}).
\end{equation}
Analogously to fully developed fluid turbulence [3], 
let us consider various moments 
of the amplitude difference between two sites.
We thus concentrate on a pair of units at sites
$\mbox{\boldmath$r$}_0$ and $\mbox{\boldmath$r$}_0+\mbox{\boldmath$x$}$
with distance $x\equiv|\mbox{\boldmath$x$}|$ satisfying $x\ll 1$.
The amplitude difference $\mbox{\boldmath$y$}_n
(\mbox{\boldmath$x$})\equiv\mbox{\boldmath$X$}_n
(\mbox{\boldmath$r$}_0+\mbox{\boldmath$x$})-
\mbox{\boldmath$X$}_n(\mbox{\boldmath$r$}_0)$  
obeys the equation
%eq.(3)
\begin{equation}
\mbox{\boldmath$y$}_{n+1}=\hat{L}_n\mbox{\boldmath$y$}_n
+O(|\mbox{\boldmath$y$}_n|^2)+\Delta\mbox{\boldmath$h$}_n,
\end{equation}
where $\hat{L}_n\mbox{\boldmath$y$}_n$ is the linearization of 
$\mbox{\boldmath$f$}(\mbox{\boldmath$X$})$ about 
$\mbox{\boldmath$X$}=\mbox{\boldmath$X$}_n
(\mbox{\boldmath$r$}_0)$, and $\Delta\mbox{\boldmath$h$}_n\equiv
\mbox{\boldmath$h$}_n(\mbox{\boldmath$r$}_0+\mbox{\boldmath$x$})-
\mbox{\boldmath$h$}_n(\mbox{\boldmath$r$}_0)$ is 
a quantity of $O(x)$.
Equation (3) describes a multiplicative stochastic process [4] with small
additive noise. Similar equations have recently aroused 
considerable interest in connection with 
on-off intermittency and related phenomena [5]. 
While on-off intermittency refers to a certain type of
temporal self-similarity peculiar to a 
special parameter value, 
our major concern below
is a {\em spatial self-similarity observable 
over an open parameter range}. \\
\indent 
Equation (3) may be simplified 
by neglecting all eigenmodes of $\hat{L}_n$
other than the least stable one. This leads to a scalar equation 
for $y_n\equiv|\mbox{\boldmath$y$}_n|$
%eq.(4)
\begin{equation}
y_{n+1}=e^{\lambda_{n}}y_n+O(y_{n}^2)+b_{n}x,
\end{equation}
where $\lambda_n$ is the local Lyapunov exponent of the unit at site 
$\mbox{\boldmath$r$}_0$, and $b_n$ is a randomly changing factor
of $O(1)$. For sufficiently small $x$, there is a range of $y$ satisfying
$x\ll y\ll 1$ where both the nonlinear and inhomogeneous terms in (4) 
are negligible. We are thus left with a linear equation 
$y_{n+1}=e^{\lambda_{n}}y_{n}$ or $z_{n+1}-z_n=\lambda_n$ in terms of
a new variable $z_n=\ln y_n$. If the random process 
of $\lambda_n$ is Markoffian, which we assume, the probability 
density $Q_n(z)$ for $z_n$ evolves in this linear regime according to
%eq.(5)
\begin{equation}
Q_{n+1}(z)=\int_{-\infty}^{\infty}w(\lambda)Q_n(z-\lambda)d\lambda,
\end{equation}
where $w(\lambda)$ is the normalized probability density for $\lambda_n$.
Equation (5) admits a stationary solution of the form 
$Q(z)\propto\exp(\beta z)\equiv y^{\beta}$. Thus, 
the corresponding probability density for $y_n$,
denoted by $P(y)$, becomes
%eq.(6)
\begin{equation}
P(y)\propto y^{-1+\beta},
\end{equation}
where $\beta$ is determined as a nontrivial (i.e., nonzero) solution of
%eq.(7)
\begin{equation}
\int_{-\infty}^{\infty}e^{-\beta\lambda}w(\lambda)d\lambda=1.
\end{equation}
Note that for sufficiently small $\beta$, we have 
%eq.(8)
\begin{equation}
\beta=2\bar{\lambda}/\bar{\lambda^2},
\end{equation}
where the bar means the average with respect to $w(\lambda)$. 
We have now to modify (6) by taking into account 
the effects of the nonlinear 
and inhomogeneous terms in (4). The nonlinearity, which is assumed 
to work in such a way that the unstable growth of $y_n$ be saturated when 
$\lambda_n>0$, may roughly be incorporated by introducing a cutoff 
in $P(y)$ at $y=1$. On the other hand, the inhomogeneous term will 
come into play when $y_n$ becomes $O(x)$ or smaller, thus suppressing 
the power-law divergence of $P(y)$ there. For the purpose of qualitative 
argument, one may therefore use the following simple model for $P(y)$:
%eq.(9)
\begin{eqnarray}
P(y)=&C&x^{-1+\beta}\;\;(y\leq x),\;\;Cy^{-1+\beta}
\;\;(x<y\leq 1),\nonumber\\
 &0&\;\;(y>1),\;\;\;C:normalization\;const.
\end{eqnarray}
This form allows us to calculate the $q$-th moment $<y(x)^q>$ for 
arbitrary $q$. For simplicity, only positive values of
$q$ will be considered below. In the subcritical regime ($\beta<0$,
i.\ e., $\bar{\lambda}<0$), where the dynamical units are entrained to 
the driving field,  
we obtain power-law moments
%eq.(10)
\begin{equation}
<y(x)^q>\sim x^{q}\;\;(q<|\beta|),\;\;
x^{|\beta|}\;\;(q>|\beta|).
\end{equation}
The result of $q$-independent exponent valid for higher 
moments ($q>|\beta|$) is anomalous reflecting strong non-Gaussianity 
of $P(y)$. In the postcritical regime ($\beta>0$), all moments possess 
an $x$-independent part, while the residual part still obeys a power law:
%eq.(11)
\begin{equation}
<y(x)^q>\sim\beta(q+\beta)^{-1}+O(x^\beta).
\end{equation} 
The reason why the amplitude difference in the postcritical regime
is nonvanishing for vanishing $x$
is that the two units in question have lost their respective
synchrony with $h_n$, implying also the loss of 
their mutual synchronization. Note that (10) and (11) are 
asymptotic formulae valid for $x\rightarrow 0$ under fixed 
$\beta$. Near $|\beta|=q$ and 0 under fixed $x$, however, there exist 
crossover regimes 
(C1) $|(\beta+q)\ln x|\ll 1$ and (C2) $|\beta\ln x|\ll 1$, respectively, 
in each of which we have 
$<y(x)^q>\sim x^{|\beta|}|\ln x|$ and $|\ln x|^{-1}$.\\
\indent 
A few more remarks are now given on the cases of $q=2$ and 1
for which our theory recovers our previous results [2]. 
We obtain from (10) 
and (11) the second moment
%eq.(12)
\begin{eqnarray}
<y(x)^2>&\sim &x^2\;\;(\beta<-2),\;\;
x^{|\beta|}\;\;(-2<\beta<0),\nonumber \\
&&\beta(2+\beta)^{-1}+O(x^\beta)\;\;(\beta>0),
\end{eqnarray}
while in the aforementioned crossover regimes, we have 
$<y(x)^2>\sim x^2|\ln x|$ (C1) and $1/|\ln x|$ (C2).\\
\indent The case $q=1$ is related to the length of an amplitude
versus space
curve. This is because the length $S(x)$ for the part of an 
amplitude profile contained in the unit interval, when measured 
with the resolution of the minimum length scale $x$, is given by 
$S(x)\sim x^{-1}<y(x)>$. Applying (10) and (11), we thus obtain 
$S(x)\sim const.$ ($\beta<-1$), $x^{|\beta|-1}$ ($-1<\beta<0$),  
and $x^{-1}$ ($\beta>0$). In the crossover regimes, however, these 
must be replaced by $S(x)\sim |\ln x|$ (C1) and $1/(x|\ln x|)$ (C2). 
The fractal dimension $D_f$ defined by $S(x)\sim x^{1-D_f}$ thus becomes
%eq.(13)
\begin{equation}
D_f=1\;\;(\beta\!<\!-1),\;\;2\!-\!|\beta|\;\;(-1\!<
\!\beta\!<\!0),\;\;2\;\;(\beta\!>\!0)
\end{equation}
except for the crossover regimes.\\
\indent The above arguments on discrete-time dynamics can easily be carried
over to continuous-time dynamics. One needs only make replacements 
$n\rightarrow t$, $n+1\rightarrow t+dt$ and
 $\lambda_n\rightarrow\lambda(t)dt$. Then, (4) becomes
$\dot y=\lambda(t)y+O(y^2)+b(t)x$, and (5) reduces to a
Fokker-Planck equation $\dot{Q}=-\bar{\lambda}\partial_{z}Q+\frac{1}{2}
\bar{\lambda^2}\partial^2_{z}Q$. The latter admits a stationary solution 
$Q(z)\propto\exp(\beta z)$, and the corresponding $P(y)$ is the same form 
as (6) with $\beta$ given by (8).\\ 
\indent 
In order to test the validity of our argument, 
the array of logistic maps 
(1) has been analyzed numerically. In Fig.\ 2, we display 
$P(y)$ versus $y$ for some values of $K$, with $x$ fixed at a sufficiently 
small value. 
As expected, $P(y)$ exhibits a power-law dependence on $y$ for not too small 
or too large $y$, with the exponent depending on $K$. 
Figure 3 shows moments $<y(x)^q>$ versus $x$ for some values of $q$. 
Their power-law dependence on $x$ is clear, but the observed change 
of the exponent with $q$, indicated in the small box, is not so sudden
across $q=|\beta|$ as the theory predicts.  As a possible source of this 
discrepancy, except for the existence of the crossover regime C1, 
the assumed form of $P(y)$ in (9) might be too simplistic, 
especially in introducing a sharp cutoff at $y=1$.\\
\indent Up to this point, we have considered uncoupled units. 
We now show some evidence that inclusion of short-range coupling 
leaves the above-described power-law behavior of moments 
essentially unchanged. 
As an illustration, we modify (1) with additional diffusive coupling:
%eq.(14)
\begin{eqnarray}
X_{n+1}(j)&=&f(X_n(j))+h_n(j)\nonumber\\
         +\frac{D}{2}\{f(X_n(j+1))&+&f(X_n(j-1))-2f(X_n(j))\}.
\end{eqnarray}
Without the forcing term $h_n$, the above model would be identical with the 
usual coupled map lattice. In Fig.\ 4, the second moments 
$<y(x)^2>$ are compared between the two systems, one with diffusive 
coupling ($D=0.1$) and the other without. The deviation from a power 
law in the presence of coupling is limited to the range covering 
ten or so units out of $N$ ($=4096$). This defines a 
lower cutoff length $x_d$ similar to the dissipation length 
in fully developed fluid turbulence. 
Although $x_d$ will increase 
with $D$ like $x_d\propto \sqrt{D}$, we have a prefactor $N^{-1}$, 
so that
$x_d$ can be made arbitrarily smaller than 1 (i.e., the upper cutoff)  
by increasing $N$ indefinitely. Thus, the intermediate range of $x$, 
which is
similar to the inertial subrange, has a sufficient extention 
over which the power-law nature of correlations
is practically unaffected. \\
\indent
It is also worth noting that whether the dynamical units 
involved are themselves chaotic or not is unimportant 
to the power-law nature of turbulent fluctuations. This is illustrated 
by driven rotators
%eq.(15)
\begin{equation}
\dot{\phi}_j=1-c\cos\phi_j+h_j(t),\;\;\;\;\;|c|<1.
\end{equation}
Obviously, each unit by itself could never be chaotic. The driving field
is assumed to be of the form
$h_j(t)=K\cos[2\pi\{\frac{j}{N} +\psi(t)\}]$,
where we let $\psi(t)$ behave like the position of a Brownian particle, 
i.e., $\dot{\psi}=av$ and $\dot{v}=-v+\xi(t)$ with $\xi(t)$ 
representing Gaussian white noise. 
Our assertion is confirmed by Fig.\ 5 which demonstrates 
power-law dependence of the second moment $<y(x)^2>$ 
on $x$ with parameter-dependent exponent. \\
\indent 
Some new aspects of our turbulent field are revealed through 
an analysis of the differential amplitudes $Y(j)$ or quantities 
defined similarly when the spatial dimension is two or higher. We call such a 
field the $Y$-field. The situation is analogous to fully developed 
fluid turbulence where the study of the energy dissipation field 
provides rich information which would hardly be available from 
the study of the velocity field alone. We will restrict our discussion 
to 1D systems below. Note that in the absence of short-range coupling
 a true derivative 
$dX/dx$ may not exist in the continuum limit, especially when the 
amplitude profile is fractal. The $Y$-field must then be redefined 
as $Y(x)\equiv\delta^{-1}|X(x+\delta)-X(x)|$ with finite but 
sufficiently small $\delta$.\\
\indent 
Spatial intermittency of $Y(x)$ as exemplified in Fig.\ 1b may be 
analyzed similarly to the case of on-off intermittency. 
This is achieved by measuring the probability density $\rho(l)$ 
for the space-interval $l$ over which the units are in the {\em laminar}
state, namely, their $Y$-values stay below 
a certain threshold $Y_0$. Such an analysis was done for the 
driven logistic maps (1). It is clear from Fig.\ 6 that, 
as in the on-off intermittency, $\rho(l)$ exhibits an inverse power law. 
We confirmed that, for not too small or too large $Y_0$, 
the exponent is insensitive to the choice of $Y_0$, but depends on $K$.\\
\indent
A more thorough characterization of the $Y$-field is provided
by the generalized fractal dimensions 
$D_q\equiv(q-1)^{-1}\lim_{\epsilon\rightarrow 0}
\ln\sum_{i}\mu_{i}^{q}/\ln\epsilon$ [6]. 
Here, the measure $\mu_{i}$ of the $i$-th box of size $\epsilon$ is 
defined as being 
proportional to the integral of $Y(x)$ within the same box, 
with the condition of the total measure being normalized. 
Similar multifractal analysis was performed for the energy dissipation 
field of fully developed fluid turbulence [7]. Figure 7 shows  
$D_q$ obtained for the driven logistic maps (1).
Note that $D_0=1$, which is simply because $Y(x)$ is nonvanishing 
almost everywhere. An open problem is to explain 
the various singularities of the 
$Y$-field and relate them to the singularities of the amplitude 
field.\\
\indent 
In conclusion, the type of turbulence discussed in this report 
is so robust that similar phenomena 
should exist quite universally. 
They may appear in a wide variety of coupled and 
uncoupled systems such as reaction-diffusion systems, fluids and 
biological populations, once placed in a long-wave randomly 
fluctuating environment.\\
\indent
The authors thank P.\ Marcq for fruitful discussions and 
careful reading of the manuscript. 
The present work has been supported by the Japanese Grant-in-Aid
for Science Research Fund from the Ministry of Education, Science 
and Culture (No.\ 07243106).

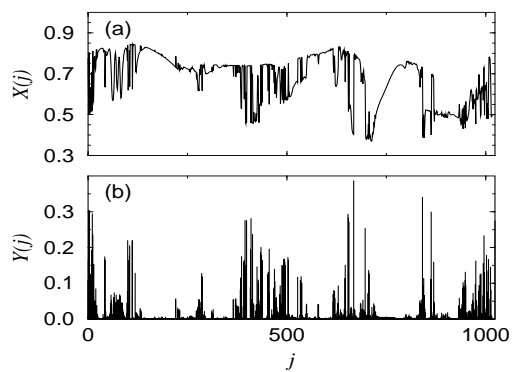
\begin{figure}
\caption{(a) Instantaneous amplitude profile for the driven logistic maps (1). 
$N=1024$, $K=0.2$, $a=3.7(1-K)$. (b) Profile of differential amplitudes
$Y(j)$ constructed from (a).}
\end{figure}
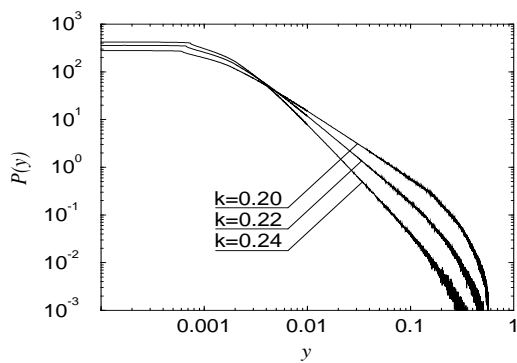
\begin{figure}
\caption{Probability density $P(y)$ in logarithmic scales for 
the driven logistic maps (1). For each $K$ value, $P(y)$ obeys 
a power-law in the intermediate range of $y$.
The exponent changes as -1.40, -1.77 and -2.33 with increasing $K$.
$a=3.7(1-K)$, $x=1024^{-1}$.} 
\end{figure}
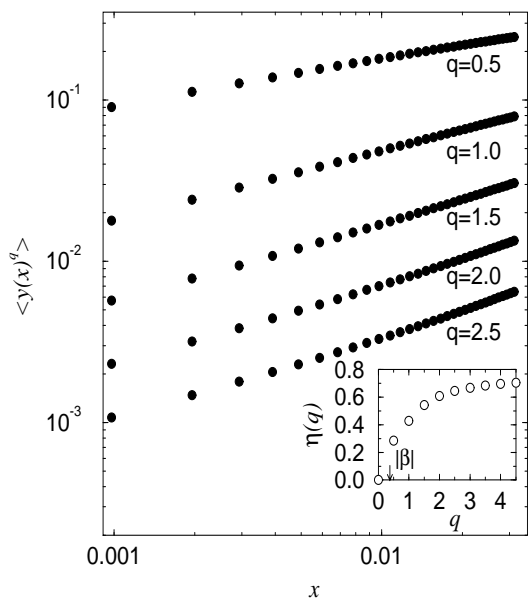
\begin{figure}
\caption{Moments $<y(x)^q>$ vs.\ $x$ in logarithmic scales
for the driven logistic maps (1), showing power-law dependence
on $x$ for each $q$ with $q$-dependent exponent $\eta(q)$.
Numerical $\eta$ vs.\ $q$ curve is displayed in the small box. 
$K=0.2$, $a=3.7(1-K)$.}
\end{figure}
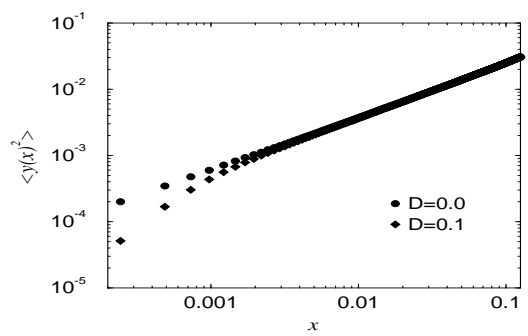
\begin{figure}
\caption{The second moment $<y(x)^2>$ vs.\ $x$ in logarithmic scales 
for the driven logistic maps (1) with additional diffusive coupling 
($D=0.1$) and without ($D=0$). $K=0.2$ $a=3.7(1-K)$, $N=4096$.}
\end{figure}
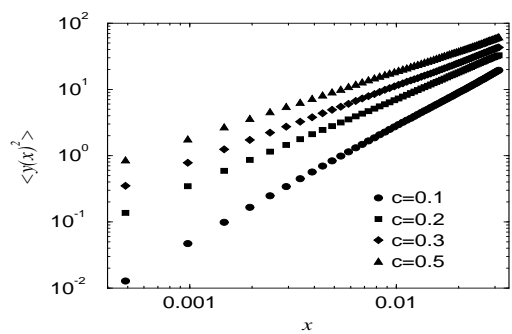
\begin{figure}
\caption{The second moments $<y(x)^2>$ vs.\ $x$ in logarithmic scales 
for the driven rotators (15) with different values of $c$. 
In numerically integrating (15), a simple Euler scheme is adopted 
where $f(t)$ is replaced with uniform random numbers over the 
interval $(-\Delta t,\Delta t]$ given at each fundamental timestep of 
$\Delta t=0.05$. $K=1.0$, $a=30.0$.}
\end{figure}
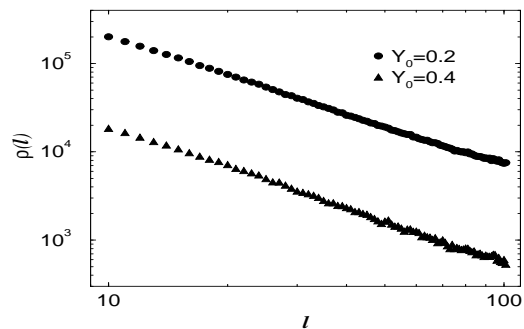
\begin{figure}
\caption{Size distribution $\rho(l)$ of the laminar domains of the 
$Y$-field ($\equiv Y(x)=|X(x+\delta)-X(x)|/\delta$) in logarithmic 
scales for the driven logistic maps (1). 
$K=0.2$, $a=3.7(1-K)$, $N=1024$, $\delta=N^{-1}$, $Y_0=0.2,$ $0.4$.}
\end{figure}
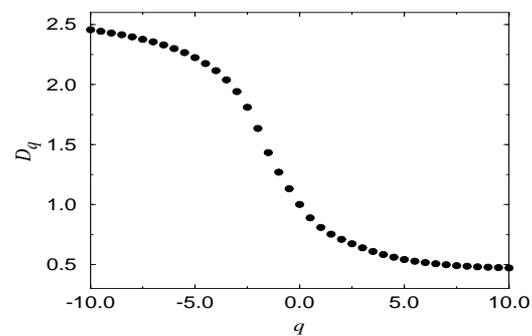
\begin{figure}
\caption{Dimension spectrum $D_q$ of the $Y$-field for the driven 
logistic maps (1). $K=0.2$, $a=3.7(1-K)$, $N=1024$, $\delta=N^{-1}$.}
\end{figure}
\end{document}